\documentclass[11pt]{article}
\usepackage{amssymb}
\usepackage[colorlinks]{hyperref}
\usepackage[dvips]{graphicx,color}

\newcommand{\ket}[1]{\left | #1 \right \rangle}
\newcommand{\bra}[1]{\left \langle #1 \right |}

\def\openone{\leavevmode\hbox{\small1\kern-3.8pt\normalsize1}}

\def\ct{{\cal T}}

\def\co{{\cal O}}

\def\cp{{\cal P}}
\def\cy{{\cal Y}}
\def\ciqp{{\cal IQP}}

\newtheorem{theorem}{Theorem}
\newtheorem{definition}{Definition}

\newtheorem{corollary}{Corollary}

\newcommand{\beq}{\begin{equation}}
\newcommand{\eeq}{\end{equation}}
\newcommand{\beqa}{\begin{eqnarray}}
\newcommand{\eeqa}{\end{eqnarray}}

\newcommand{\poly}{{\rm poly}}
\newcommand{\prob}{{\rm prob}\,}

\begin{document}
\begin{center}
{\LARGE\bf Classical simulation of commuting quantum computations implies collapse of the polynomial hierarchy}\\
\bigskip
{\normalsize Michael J. Bremner$^1$, Richard Jozsa$^2$ and Dan J. Shepherd$^3$}\\
\bigskip
{\small\it $^1$Institut f\"ur Theoretische Physik, Leibniz Universit\"at Hannover,\\
Appelstrasse 2, 30167 Hannover\\[1mm]
$^2$DAMTP, Centre for Mathematical Sciences, University of Cambridge,\\ Wilberforce Road, Cambridge CB3 0WA, U.K.\\[1mm]
$^3$CESG, Hubble Road, Cheltenham, GL51 0EX, U.K.\\
}
\end{center}

\begin{abstract} We consider quantum computations comprising only commuting gates, known as IQP computations, and provide compelling evidence that the task of sampling their output probability distributions is unlikely to be achievable by any efficient classical means. More specifically we introduce the class post-IQP of languages decided with bounded error by uniform families of IQP circuits with post-selection, and prove first that post-IQP equals the classical class PP. Using this result we show that if the output distributions of uniform IQP circuit families could be classically efficiently sampled, even up to 41\% multiplicative error in the probabilities, then the infinite tower of classical complexity classes known as the polynomial hierarchy, would collapse to its third level. We mention some further results on the classical simulation properties of IQP circuit families, in particular showing that if the output distribution results from measurements on only O(log n) lines then it may in fact, be classically efficiently sampled.
\end{abstract}
\bigskip

\section{Introduction}\label{intro} 
From a pragmatic point of view the field
of quantum computing is driven by the expectation that quantum algorithms can
offer some computational complexity benefits transcending the possibilities of
classical computing. But this expectation can be challenged both theoretically and experimentally: (a) there is yet no theoretical proof that any quantum algorithm outperforms the best classical algorithm for the task, in the standard computational setting of polynomial vs. exponential running time (without inclusion of further constructs, such as use of oracles, or consideration of distributed computing and the role of communication; in both these scenarios there are indeed proofs of exponential complexity benefits); (b) experimentally there are well documented difficulties associated with building a quantum computer that is suitably fault tolerant and sufficiently scalable to manifestly demonstrate a complexity benefit.

However both (a) and (b) can, to some extent, be redressed by further examination: the criticism in (a) can be attributed to limitations of {\em classical} complexity theory -- we do have interesting quantum algorithms (such as Shor's factoring algorithm) for problems widely believed to be classically hard but there is no proof of the latter. Proof of classical hardness is a notoriously difficult issue (cf the famous P vs. NP question) and it has become popular to resort to providing only evidence of hardness, as follows: we prove that if a certain problem were classically easy then this would entail consequences that are highly implausible (although also generally unproven) e.g. collapse of an entire complexity class (such as entailing that P $=$ NP). For (b) we could seek to devise a computational task that, on the one hand is expected to be classically hard (as above) yet on the other hand, can be implemented using suitably simple (sub-universal) quantum computational elements that are especially easily or fault-tolerantly implementable within some specific experimental scheme.
In this paper we develop a family of such computational tasks (that amount to sampling from suitably prescribed probability distributions). Recently a different approach to similar issues has been described by Aaronson in \cite{qiptalk}. More generally there has been increasing interest in physically restricted forms of quantum computing and a study of associated complexity classes \cite{bkashper,jordan,shbr,maarten,jkmw}.

We consider so-called temporally unstructured quantum computations (also known as IQP or ``instantaneous'' quantum computation) introduced in \cite{danthesis,shbr}. Our main result is to demonstrate that if quantum circuits comprising 2-qubit {\em commuting} gates could be simulated classically (even up to a generous multiplicative error tolerance as described below) then the infinite tower of complexity classes known as the polynomial hierarchy (PH), would collapse to its third level. While not implying that P$=$NP, such a collapse is nevertheless widely regarded as being similarly implausible. Apart from their tantalising theoretical simplicity, such circuits of only commuting gates are known to be of significance for super- and semi-conductor qubit implementations, where it has recently been shown \cite{alif}  that they are much simpler to implement fault-tolerantly than gates drawn from a fully universal set.

A significant ingredient in our derivations will be the notion of a post-selected quantum computation. Aaronson \cite{aarpost} has shown that if post-selection is included with universal polynomial time quantum computation then the computational power is boosted from BQP to the classical class PP. We will show that, somewhat surprisingly, post-selection boosts the power of the much weaker class of polynomial time IQP computations to PP too.

The notion of classical simulation that applies in our main result is an especially weak one -- broadly speaking (cf precise definitions below) given a description of a quantum circuit we ask for a classical process that can provide a {\em sample} of a probability distribution, that approximates the output distribution of the quantum process to a suitable multiplicative accuracy. A very much stronger notion of simulation sometimes used in the literature (which we shall call a strong simulation) is to ask for a classical efficient computation of the {\em value} of any marginal or total output probability, to exponential precision.  Previously it was known \cite{terdiv,fenetc} that the existence of such strong simulations for some classes of quantum computations would imply the collapse of the polynomial hierarchy. Our result contrasts with these works in the striking simplicity of the quantum processes being considered and in the very much weaker requirements in the classical simulation.

\section{Preliminary notions}\label{prelim}
We begin by introducing some definitions and notations needed to give precise statements of our main results.
\subsection{Computational tasks}\label{taskdef}
Conventionally a computational task $\ct$ is a specified relationship between inputs $w=x_1\ldots x_n$ and outputs $ y_1\ldots y_m = \ct (w)$ which are taken to be bit strings. The length $n$ of the input string is called the input size. A computational process $C$ with (generally  probabilistic) output $C(w)$ on input $w$ is said to compute $\ct$ with {\em bounded error} if there is a constant $0 < \epsilon < \frac{1}{2}$ such that for all inputs, $\prob [C(w)=\ct (w)] \geq 1-\epsilon$. $C$ computes $\ct$ with {\em unbounded error} if for all inputs we have $\prob [C(w)=\ct (w)] > \frac{1}{2}$. If the output of $\ct$ is always a single bit then $\ct$ is called a decision task associated to the subset $\{ w: \ct (w)=1 \}$ of all bit strings. A subset of bit strings is called a language.

A more general kind of computational task involves merely the sampling of a probability distribution on $m$-bit strings whose result is not necessarily associated to any desired ``correct'' outcome $j_1 \ldots j_m$ as above. For example, for each $n$-bit string $w$ we may have an associated quantum circuit $C_w$ with output probability distribution $P_w$ on $m$-bit strings, and we may be interested to know how hard it is to sample from $P_w$ by purely classical means, given a description of the circuit $C_w$.

\subsection{Uniform families of circuits}\label{unifcir}
We shall use a notion of uniform circuit family that is slightly different from the standard textbook definition, motivated by a desire to make more transparent the contribution of the uniformity condition to the final overall computational process.

In the Turing machine model of computation a single machine suffices to deal with inputs of all sizes. In contrast in the circuit model, any single circuit has a fixed number of input lines so to treat inputs of all sizes it is conventional to introduce the notion of a circuit {\em family} $\{ C_n \} = \{ C_1,C_2, \ldots \}$ with $C_n$ being a circuit intended to perform the computation for all inputs of size $n$. In this formalism we need to impose an auxiliary {\em uniformity condition} specifying computational limitations (cf below) on how the (descriptions of the) circuits $C_n$ themselves are generated as a function of $n$. In the absence of any such condition, hard (or even uncomputable) computational results may, by fiat, be hard wired into the varying structure of the circuits $C_n$ with $n$. In standard treatments a circuit family $\{ C_n \}$ is parameterised by input size $n$ (with $C_n$ being a circuit processing all inputs of size $n$). For our purposes it will be more convenient to parameterise the circuit family by the inputs $w=x_1\ldots x_n$ themselves, with circuits always acting on a standard input such as $0\ldots 0$ (or $\ket{0}\ldots \ket{0}$ for quantum circuits), resulting in circuit families denoted $\{ C_w \}$. Thus for example in comparison with the standard definition, we could take the circuit $C_w$ to be the circuit $C_n$ prefixed by some NOT gates (depending on $w$) that initially convert the input $0\ldots 0$ into $w$. Our formal definition is as follows.

\begin{definition}\label{cirfamdef} 
A uniform family of circuits (of some specified type) is a mapping $w \rightarrow C_w$ where $w=x_1\ldots x_n$ is a bit string of length $n$, $C_w$ is a (classical) description of a circuit (of the appropriate type) and the mapping $w\rightarrow C_w$ is computable in classical $\poly(n)$ time. Here the description $C_w$ includes (i) a specification of a sequence of gates and lines upon which they act, (ii) a specification of the inputs for all lines (often taken to be $0\ldots 0$ resp. $\ket{0}\ldots \ket{0}$ for classical resp. quantum circuits), (iii) a specification of which lines comprise the output register, and (iv) a specification of any other registers needed for a circuit of the type being used (e.g. a register of lines initialised to random bit values for randomised computation, or a register of post-selection lines for post-selected computations, as defined later).

Associated to any uniform circuit family we have a family of probability distributions $\{ P_w \}$ (on $m$-bit strings where $m$ is the size of the output register of $C_w$), defined by the output of the computational process described by $C_w$.
\end{definition}
Since $w\rightarrow C_w$ is computable in $\poly (n)$ time, each circuit $C_w$ has $\poly(n)$ size and acts on at most $\poly(n)$ lines. One may entertain other uniformity conditions e.g. having $w\rightarrow C_w$ computable in classical log space (as is generally adopted for $n\rightarrow C_n$ in the textbook definition of uniform families). For us the $\poly(n)$ time uniformity condition is adequate, as we are primarily interested in circuits whose computational power is potentially stronger than, or not commensurate with, classical deterministic polynomial time. Our uniformity definition (based on inputs $w$ rather than just input sizes $n$) then transparently simply prefixes the processing power of the circuits with arbitrary classical deterministic polynomial time computation.

For classical deterministic polynomial time computation in our circuit family definition, the computation can be totally represented within the uniformity stage $w\rightarrow C_w$  and the $C_w$'s can be taken to be trivial circuits that perform no further computation beyond outputting the obtained answer. Classical {\em randomised} polynomial time computation is modelled by circuits $C_w$ that have a designated register of lines (disjoint from the input register) which is initialised with random bits for each run of the computation $C_w$. Such circuits are called classical randomised circuits. The complexity class of decision tasks decided with bounded error (resp. unbounded error) by uniform families of classical randomised circuits is denoted BPP (resp. PP). It is well known that BPP is independent of the value of the constant error tolerance $\epsilon$. For universal polynomial time {\em quantum} computation the circuits $C_w$ comprise quantum gates, each acting on a constant number of lines. The input is taken to be the standard state $\ket{0}\ldots \ket{0}$  and the output is the (probabilistic) result of a computational basis measurement on a designated register of output lines. The class of decision tasks solved with bounded error by such uniform families is denoted BQP. (This definition is easily seen to be equivalent to other standard definitions of BQP such as in \cite{nc}).

\subsection{IQP circuits}
We now come to our notion of quantum computations comprising commuting gates. In \cite{shbr} these have been called IQP  (``instantaneous quantum polynomial time'') computations since in quantum physics, such gates may be applied simultaneously.

\begin{definition}\label{defiqp}  
An IQP circuit on $n$ qubit lines is a quantum circuit with the following structure: each gate in the circuit is diagonal in the $X$ basis $\{ \ket{0}\pm \ket{1} \}$, the input state is $\ket{0}\ket{0}\ldots \ket{0}$ and the output is the result of a computational basis measurement on a specified set of output lines.
\end{definition}
In this paper we will assume that each gate in the description of an IQP circuit $C_w$ is specified by giving its diagonal entries and the lines on which it acts. Thus a $\poly (n)$ sized description implies that any gate acts on at most $O(\log n)$ lines. We note however that other inequivalent conventions are possible e.g. in \cite{shbr} gates are specified by a parameter $\theta$ and a subset $i_1,\ldots , i_k$ of lines, corresponding to the gate $U=\exp (i\theta X_{i_1}\otimes \ldots \otimes X_{i_k})$ which may thus act on $O(n)$ lines, although its (potentially exponentially many) diagonal entry phases $\pm \theta$ are all equal up to sign.

It will sometimes be convenient to represent an IQP circuit in terms of gates diagonal in the $Z$ (or computational) basis. In this representation the inputs and outputs are the same as before but the circuit of gates is required to have the following structure: each qubit line begins and ends with a Hadamard ($H$) gate, and in between, every gate is diagonal in the $Z$ basis. This is easily seen to be equivalent to the previous definition (by inserting two $H$'s on each line between each pair of gates, recalling that $HH=I$, and then absorbing all $H$'s into conjugation actions on the $Z$ basis diagonal gates, leaving only $X$ basis diagonal gates).

As noted in definition \ref{cirfamdef} any uniform circuit family $\{ C_w \}$ associates a probability distribution $P_w$ to each bit string $w$ and we will be especially interested to consider whether this distribution can be sampled (to suitable accuracy) by purely classical means in $\poly (n)$ time, given the classical description of the circuit $C_w$. For this issue it will be significant to note the {\em number} of output lines, and especially its growth with $n$.

\subsection{Post-selected circuits}
An important theoretical tool in our arguments will be the notion of a {\em post-selected} (classical or quantum) circuit $C$. This is a circuit which, in addition to a specified register of output lines $\co$, has a further (disjoint) specified register of post-selection lines $\cp$. Then instead of sampling measurement results $x$ directly from the output lines with distribution $\prob [ \co =x]$, we consider only those runs of the process for which a measurement on the post-selection lines yields $00\ldots 0$ i.e. the output distribution on $\co$ is now taken to be the conditional distribution $\prob [\co = x| \cp = 00\ldots 0]$. In this construction we also require the circuit $C$ to have the property that $\prob [\cp = 00\ldots 0] \neq 0$ so that the conditional probabilities are well defined:
\begin{equation}\label{bayes} \prob [\co = x| \cp = 00\ldots 0] = \frac{\prob [\co = x \,\&\, \cp = 00\ldots 0]}{\prob [\cp = 00\ldots 0]}.
\end{equation}
In practical terms a post-selected computation would be implemented by repeatedly running the computation and considering the output register only if the post-selection register is measured to yield $00\ldots 0$. Since we place no limit on how small the (non-zero) probability of the latter event may be, the post-selection process may incur an exponential overhead in time, and similar to the notion of a non-deterministic computation, it is principally of interest as a theoretical tool rather than as a feasible computational resource.

\begin{definition}\label{postiqp} 
A language $L$ is in the class post-IQP (resp. post-BQP or post-BPP) iff there is an error tolerance $0<\epsilon < \frac{1}{2}$ and a uniform family $\{ C_w \}$ of post-selected IQP (resp. quantum or randomised classical) circuits with a specified single line output register $\co_w$ (for the $L$-membership decision problem) and a specified (generally $O(\poly (n))$-line) post-selection register $\cp_w$ such that:\\ (i) if $w\in L$ then $\prob [\co_w = 1| \cp_w = 00\ldots 0]\geq 1-\epsilon$ and\\
(ii) if $w\notin L$ then $\prob [\co_w = 0| \cp_w = 00\ldots 0]\geq 1-\epsilon$.
\end{definition}
It is pertinent to remark on the $\epsilon$-independence of the classes in definition \ref{postiqp} above. The basic bounded error classes BPP and BQP are well known to be independent of the error tolerance $0<\epsilon <\frac{1}{2}$. Indeed the standard method  \cite{ab,papadim} for reducing $\epsilon$ is to consider the majority vote answer of multiple runs of the circuit. Similarly post-BPP and post-BQP are easily seen to be independent of the error tolerance value too. The class post-IQP is in fact also independent of $\epsilon$, as will follow from theorem \ref{postthm} below. However the class BIQP$_\epsilon$ of languages decided with bounded error $\epsilon$ by uniform families of IQP circuits (with no post-selection) is not known to be independent of $\epsilon$ as it is not evident whether or not the majority vote function can be realised by just (commuting) IQP circuits. Fortunately we will not need to directly use BIQP$_\epsilon$ in our arguments.

Post-selected classical computation has been considered in \cite{kuper,hanhema}. The class called BPP$_{\rm path}$ that is extensively studied in \cite{hanhema} is easily seen to be equal to our class post-BPP.

For quantum computation, the class post-BQP was introduced and studied by Aaronson in \cite{aarpost} where is was shown that post-BQP equals the classical class PP. Note that if general quantum or classical circuits are available, it suffices (as in \cite{aarpost}) to use post-selection registers of only a {\em single} line, since for any register of $k$ lines we may adjoin a circuit that computes some simple function $f$ with $f(x_1\ldots x_k) = 0$ iff $x_1\ldots x_k = 00\ldots 0$ e.g. the OR of the $k$ bit values suffices. However if the allowed gates are restricted (as in the case of IQP circuits) it may not be possible to compute any such function using only the allowed resources, and post-selection on multiple lines needs to be entertained, as in our definition above.

\subsection{Notions of classical simulation for quantum circuits}
There are various possible notions of classical simulation for quantum circuit families.
For any uniform family $\{ C_w \}$ let $P_w$ denote the
output distribution of $C_w$ and let $n$ denote the length of $w$.

\noindent
(a) We say that a circuit family
is {\em strongly simulable} if any output probability in $P_w$ and any marginal probability of $P_w$
can be computed to $m$ digits of
precision  in classical poly$(n,m)$ time.

\noindent
(b) A circuit family is {\em weakly simulable} if
given the description of $C_w$, its output distribution $P_w$ can be sampled by purely classical means
in poly$(n)$ time. Note that strong simulability implies weak simulability \cite{terdiv} -- although the sample space of $P_w$ is exponentially large in $n$ we can sample the distribution in $\poly (n)$ time by successively sampling the bits; the binary distribution used for each successive bit is the conditional distribution, conditioned on the already seen values, and these two conditional probabilities can be computed in $\poly (n)$ time via Bayes' rule, as a quotient of two marginal probabilities of $P_w$.

Next we have some notions of approximate classical simulation.

\noindent
(c) A circuit family is {\em weakly simulable with multiplicative error $c\geq 1$} if there is
a family $R_w$ of distributions (on the same sample spaces as $P_w$) such that $R_w$ can be sampled in
classical poly$(n)$ time and for all $x$ and $w$ we have
\begin{equation}\label{multsim}  
\frac{1}{c}\,\, \prob [P_w = x] \leq \prob [R_w =x] \leq  c\,\, \prob [P_w =x].
\end{equation}
(d) A circuit family is {\em weakly simulable within $\epsilon$ total variation distance} if there is a family $R_w$ as in (c) above, but with eq. (\ref{multsim}) replaced by the condition
\[ \sum_x \left| \prob [P_w=x]-\prob [R_w = x]\right| <\epsilon .\]
(e) A further notion of
approximate weak simulation has been formulated in \cite{maarten}: recall first that the Chernoff-Hoeffding
bound (cf Appendix of \cite{maarten}) implies the following result -- if we have a quantum process implementing $C_w$ then by running it poly-many times we can (with probability exponentially
close to 1) obtain an estimate $\tilde{p}$ of any output probability $p$ to within polynomial precision i.e
for any polynomial $f(n)$ we can output $\tilde{p}$ such that $|p-\tilde{p}| < 1/f(n)$. We say that
a circuit family is (classically) {\em weakly simulatable with additive polynomial error} if the
same estimates can be obtained from
the circuit descriptions $C_w$ by purely classical means in poly$(n)$ time (and probability
exponentially close to 1). Thus weak simulability implies weak simulability with additive polynomial error.

Note that if a uniform circuit family $C_w$ decides a
language $L$ with bounded error probability $0 < \epsilon <\frac{1}{2}$
then the existence of a weak (resp. strong) simulation for $C_w$ implies
that $L \in$ BPP (resp. P). Similarly the existence of a weak simulation with additive polynomial error, or with
multiplicative error $1 \leq c< 2(1-\epsilon )$, will also imply that $L \in$ BPP. The latter condition on $c$ serves to guarantee that $R_w$ still decides $L$ with a bounded error $0<\epsilon' <\frac{1}{2}$.

\section{Main results}\label{results}
\subsection{The power of IQP with post-selection}\label{pspower}
We begin by examining how the availability of post-selection is able to boost the computational power of various classes of circuits. For this it is convenient to introduce some further notions from complexity theory. If $A$ and $B$ are complexity classes, $A^B$ denotes the class $A$ with an oracle for $B$ (cf \cite{ab,papadim} for formal definitions). We may think of $A^B$ as the class of languages decided by the computations subject to the restrictions and acceptance criteria of $A$ but allowing an extra new kind of computational step: we have an oracle or ``subroutine'' for any desired language $L$ in $B$ that may be queried at any stage in the course of the computation, and each such query counts as a single computational step i.e. bit strings may be generated as intermediate results and presented to the oracle, which in a single step, returns the information of whether the bit string is in $L$ or not. The polynomial hierarchy class PH \cite{ab,papadim} is defined to be the union of an infinite tower of increasing classes $\Delta_k$, $k=1,2,\ldots$, in which $\Delta_1=$ P and $\Delta_{k+1}=$ P$^{\rm{N}\Delta_k}$. 
Here N$\Delta_k$ denotes the non-deterministic class associated to $\Delta_k$, in the same way that NP denotes the non-deterministic class associated to P, i.e.\@ we allow the process to branch at each step into two separate computational paths and deem it to accept its input if and only if at least one path accepts.
Further discussion and alternative characterisations of PH may be found in \cite{ab,papadim}.

For classical computation it is known \cite{ab,papadim} that BPP is contained in N$\Delta_2$ and also that post-BPP is contained in $\Delta_3$ \cite{hanhema}.
Now for any complexity class C we have ${\rm P}^{({\rm P}^{\rm C})} = {\rm P}^{\rm C}$ (since in the first expression any query to a ${\rm P}^{\rm C}$ oracle can be replaced by a polynomial time computation with queries to the corresponding oracle for C). Hence we get
\begin{equation}\label{ph3} 
{\rm P}^{\rm post-BPP}\subseteq {\rm P}^{\Delta_3} = \Delta_3. 
\end{equation}
We will use this inclusion below in corollary \ref{maincor}.

For the case of quantum computation it is not known whether BQP is contained within PH or not \cite{aarph}, but as mentioned above, Aaronson\cite{aarpost} has shown that post-BQP$=$PP. A theorem of Toda \cite{toda,ab,papadim} asserts that PH$\subseteq$P$^{\rm PP}$ so we get ${\rm P}^{\rm post-BQP} \supseteq {\rm PH}$. On the other hand we had  ${\rm P}^{\rm post-BPP}\subseteq \Delta_3$ so from an oracle perspective, the power of post-BPP is modest compared to post-BQP or PH.

In view of the above considerations, and recalling that uniform families of IQP circuits are intuitively expected to be far weaker than general quantum computations (and even fail to include many computations in P, such as many elementary arithmetic operations that manifestly depend on the order of operations applied) our next result is perhaps unexpected.
\begin{theorem}\label{postthm} 
post-IQP $=$ post-BQP $=$ PP.
\end{theorem}
{\bf Proof.} Clearly post-IQP $\subseteq$ post-BQP and we show the reverse inclusion. Consider an arbitrary uniform quantum circuit family with inputs $\ket{0}\ldots \ket{0}$ and with gates drawn from the following universal set: $H,Z,CZ$ and $P=e^{i\frac{\pi}{8}Z}$. (For a later purpose we point out here that all these gates are 1- or 2-qubit gates and apart from $H$, all gates have diagonal entries that are integer powers of $e^{i\pi /8}$.) If we are allowed to post-select such circuit families then we obtain post-BQP as the class of languages decideable with bounded error. Our strategy is to exhibit a direct reduction from any such post-selected circuit family to a post-selected IQP circuit family whose output conditional probabilities are the same as those of the original family.

Firstly we add in extra $H$ gates to ensure that every line begins and ends with an $H$ gate. This is possible since $H^2=I$. Next consider in turn each intermediate $H$ gate i.e. those that do not begin or end a line. For each such gate $H_a$ acting on line $a$ we include an extra qubit line labelled $e$ (for ``extra''). Consider now the following ``Hadamard gadget'' (somewhat akin to a gate teleportation) illustrated in figure 1. On lines $ae$ initialised to $\ket{\psi}_a\ket{0}_e$, where $\ket{\psi}$ is any state, we apply the process $\ket{\psi}_a\ket{0}_e \rightarrow H_a CZ_{ae}H_e  \ket{\psi}_a\ket{0}_e$ followed by post-selection of outcome 0 on line $a$. An easy calculation shows that the resulting state on line $e$ is $H\ket{\psi}$. In the original circuit we replace $H_a$ by the Hadamard gadget; here $\ket{\psi}$ represents the circuit's general input state to $H_a$ and subsequently line $e$ is used as the output line of $H_a$ for further gates in the original circuit. Alternatively we may extend the gadget by a SWAP$_{ae}$ gate and use line $a$ as output. SWAP is not a valid IQP gate so to obtain the final circuit we commute out all SWAP gates to the end of the lines.

In the resulting circuit, the new line $e$ is initialised to $\ket{0}$ and begins and ends with an $H$ gate. Thus the non-diagonal intermediate $H$ gate has been replaced by a new $CZ$ gate and an additional post-selection. Performing this replacement for every intermediate $H$ gate results in an IQP circuit with some extra post-selections on the new $e$ lines, and with the same output conditional probabilities as originally (now conditioned on the new extra post-selections too). $\Box$

\begin{figure}[!ht]
\begin{center}

  \setlength{\unitlength}{1cm}
  \begin{picture}(14,7)
  
  \put(0.2,5.4){\bf (a)}
  
  \put(1.6,5.5){$\ldots$}  \put(2.3,5.4){$\ket{\alpha}$}
  \put(4,5){\framebox(1,1){$U$}}
  \put(6,5){\framebox(1,1){$H$}}
  \put(8,5){\framebox(1,1){$V$}}
  \put(3,5.5){\line(1,0){1}}  \put(5,5.5){\line(1,0){1}}  \put(7,5.5){\line(1,0){1}}  \put(9,5.5){\line(1,0){1}}
  \put(10.2,5.4){$VHU\ket{\alpha}$}  \put(12,5.5){$\ldots$}
  
  \put(0.2,2.3){\bf (b)}

  \put(1.6,3.3){$\ldots$}  \put(2.3,3.2){$\ket{\alpha}$}
  \put(4,2.8){\framebox(1,1){$U$}}
  \put(7.5,2.8){\framebox(1,1){$H$}}
  \put(3,3.3){\line(1,0){1}}  \put(5,3.3){\line(1,0){2.5}}  \put(8.5,3.3){\line(1,0){0.5}}
  \put(9.2,3.2){$\bra{0}$}
  
  \put(6.2,2.25){$CZ$}
  \put(6.5,1.9){\line(1,1){0.5}}  \put(6.5,1.9){\line(-1,1){0.5}}  \put(6.5,2.9){\line(1,-1){0.5}}  \put(6.5,2.9){\line(-1,-1){0.5}}
  \put(6.5,3.3){\circle*{0.2}}  \put(6.5,1.5){\circle*{0.2}}
  \put(6.5,2.9){\line(0,1){0.4}}  \put(6.5,1.5){\line(0,1){0.4}}

  \put(3.4,1.4){$\ket{0}$}  
  \put(4.5,1){\framebox(1,1){$H$}}
  \put(8,1){\framebox(1,1){$V$}}
  \put(4,1.5){\line(1,0){0.5}}  \put(5.5,1.5){\line(1,0){2.5}}  \put(9,1.5){\line(1,0){1}}
  \put(10.2,1.4){$VHU\ket{\alpha}$}  \put(12,1.5){$\ldots$}

  \end{picture}
\end{center}
\noindent {\bf Figure 1}: The Hadamard gadget for removal of intermediate $H$ gates. {\bf (a)} $\ket{\alpha}$ represents a general input state to a gate $U$ within the circuit that is followed by an intermediate $H$ gate. {\bf (b)} The lower line is a new ancillary qubit line. The original intermediate $H$ gate may then be replaced by a new $CZ$ gate, a post-selection (denoted by $\bra{0}$) and two $H$ gates that are now both at the ends of lines, as allowed in IQP circuit architecture.

\end{figure}

In the above construction, the post-BQP circuit that we start with, may without loss of generality, be assumed to comprise only nearest-neighbour 2-qubit gates.  Then the SWAP operations introduced by the Hadamard gadgets will at first sight, result in a post-IQP circuit that is not truly nearest-neighbour. But by simply  `terminating some of the measurements early', and `creating some ancillas late' we can avoid line crossings (as is evident from figure 1(b)).  The practical upshot of this is that the quantum part of the IQP process resulting from this construction can be rendered, logically speaking, by local interactions on a flat 2-dimensional surface (albeit still involving the inefficient resource of post-selection).

\subsection{Classical simulation of IQP circuits and collapse of PH}
Although IQP circuits have very simple ingredients, we now provide evidence (in corollary~\ref{maincor} below) that they nevertheless embody computational possibilities that are inaccessible to classical efficient (randomised) computation.
\begin{theorem}\label{mainthm} If the output probability distributions generated by uniform families of IQP circuits could be weakly classically simulated to within multiplicative error $1\le c<\sqrt{2}$ then post-BPP $=$ PP.\end{theorem}
{\bf Proof.} We will show that under the stated simulation assumption, any language in post-IQP is in post-BPP and then theorem \ref{postthm} (together with post-BPP $\subseteq$ post-BQP) will give post-BPP $=$ PP.

Let $L \in $post-IQP be any language decided with bounded error by a uniform family of post-selected IQP circuits $C_w$ with (single line) output registers $\co_w$ and postselection registers $\cp_w$. Introduce
\begin{equation}\label{squot} S_w(x) = \frac{\prob [\co_w=x \,\&\, \cp_w = 0\ldots 0]}{\prob [ \cp_w = 0\ldots 0]} \end{equation}
so the bounded error condition states the following:
\begin{equation}\label{bdederr} \begin{array}{l}
\mbox{if $w\in L$ then $S_w(1)\geq \frac{1}{2}+\delta$} \\
\mbox{if $w\notin L$ then $S_w(1)\leq \frac{1}{2}-\delta$}
\end{array}  \end{equation}
for some $0<\delta <\frac{1}{2}$. Furthermore post-IQP is independent of the level of error so for any $L\in$ post-IQP we may assume that eq. (\ref{bdederr}) holds for any choice of $0<\delta <\frac{1}{2}$, however large. Now let $\cy_w$ denote the full register of lines of $C_w$, comprising $m$ lines say. If an output measurement on all lines of $C_w$ can be weakly classically simulated to within multiplicative error $c$ then there is a uniform family of classical randomised circuits $\tilde{C}_w$ with output register $\tilde{\cy}_w$ comprising $m$ lines with
\begin{equation}\label{multapp} \frac{1}{c}\,\prob [ \cy_w = y_1\ldots y_m] \leq \prob [ \tilde{\cy}_w = y_1\ldots y_m] \leq c\, \prob [ \cy_w = y_1\ldots y_m]. \end{equation}
Similarly all marginal distributions for corresponding sub-registers of $\cy_w$ and $\tilde{\cy}_w$ satisfy the same inequality. Let $\tilde{\co}_w$ and $\tilde{\cp}_w$ denote the registers of $\tilde{C}_w$ corresponding to $\co_w$ and $\cp_w$ of $C_w$, and introduce
\begin{equation}\label{stwiddle} \tilde{S}_w(x) = \frac{\prob [\tilde{\co}_w=x \,\&\, \tilde{\cp}_w = 0\ldots 0]}{\prob [ \tilde{\cp}_w = 0\ldots 0]}. \end{equation}
Using the inequalities of eq. (\ref{multapp}) (for the registers appearing in eq. (\ref{stwiddle})) we get
\begin{equation}\label{gocompare} \frac{1}{c^2} S_w(x) \leq \tilde{S}_w(x) \leq c^2 S_w(x). \end{equation}
Combining this with eq. (\ref{bdederr}) we see that the classical uniform family $\tilde{C}_w$ (post-selected on $\tilde{\cp}_w$) will decide $L$ with bounded error if $c^2 <1+2\delta$. Since $\delta$ can be any value satisfying $\delta <\frac{1}{2}$ we see that any value of $c<\sqrt{2}$ will suffice to guarantee that $L\in$ post-BPP.\, $\Box$

It is interesting to point out that our use of a multiplicative approximation (cf eq. (\ref{multapp})) accords well with the quotient structure of the conditional probabilities in eqs. (\ref{squot}) and (\ref{stwiddle}), allowing us to derive the bounding relationship eq. (\ref{gocompare}) between $S_w$ and $\tilde{S}_w$. In contrast, use of an additive approximation or approximation to within $\epsilon$ total variation distance would be problematic: the denominators of eqs. (\ref{squot}) and (\ref{stwiddle}) are required only to be positive, so additive or total variation distance approximations would allow catastrophic divergences of the associated probability quotients.

\begin{corollary}\label{maincor} If the output probability distributions generated by uniform families of IQP circuits could be weakly classically simulated to within multiplicative error $1\le c<\sqrt{2}$ then the polynomial hierarchy would collapse to its third level i.e. PH $= \Delta_3$. \end{corollary}
{\bf Proof.} Under the simulation assumption we may apply theorem \ref{mainthm}, and Toda's theorem with eq. (\ref{ph3}) gives ${\rm PH} \subseteq {\rm P}^{\rm PP} = {\rm P}^{\rm post-BPP} \subseteq \Delta_3$.\, $\Box$ 

From the proof of theorem \ref{postthm} we see that it suffices in theorem \ref{mainthm} and corollary \ref{maincor} to require the weak simulability condition only for a restricted kind of IQP circuit family, namely those comprising only 1- and 2-qubit gates with diagonal entries being only integer powers of $e^{i\pi /8}$.
In a similar vein one may ask whether the output register may be able to be restricted too, e.g. to having size only $O(\log n)$. Recall that for the class post-IQP, although we have only single-line output registers, the post-selection register may generally have size $O(\poly (n))$ and in the proof of theorem \ref{mainthm}, the classical simulation needs to be applicable to IQP circuit families whith output registers of the latter size too (as they incorporate the original post-selection registers). Our next result shows that such restriction on the size of the output or post-selection register is not possible (on the assumption that PH does not collapse) i.e. we see that the computational power of post-selected IQP circuits (with a single line output register) depends crucially on the size of the post-selection register.
\begin{theorem}\label{simlogthm} Let $P_w$ be the output probability distributions for any uniform family of IQP circuits in which the output registers have size $O(\log n)$. Then $P_w$ may be sampled (without approximation) by a classical randomised process that runs in time $O(\poly (n))$.\end{theorem}
{\bf Proof.}  Let $C_w$ be any uniform family of IQP circuits with output registers $\co_w$ of size $M=O(\log n)$. Let $\cy_w$, of size $N$, denote the complementary register of all non-output lines and let $x$ and $y$ denote generic bit strings of lengths $M$ and $N$ respectively. We view $C_w$ in its $Z$-basis diagonal representation: on input $\ket{0}\ldots \ket{0}$ the initial Hadamard gates on all lines create an equal superposition and after all $Z$-diagonal gates of the circuit (and just before the final round of Hadamard gates) the state has the form
\begin{equation}\label{state} \ket{\phi}=\frac{1}{\sqrt{2^{M+N}}} \sum_{x,y} e^{if(x,y)} \ket{x,y}. \end{equation}
The phase function $f(x,y)$ can be computed in classical $\poly (n)$ time by accumulating the relevant diagonal elements of the successive gates. Now the result of further gates and measurements on $\co_w$ is independent of measurements on the disjoint register $\cy_w$. According to eq. (\ref{state}) a measurement of $\cy_w$ will yield a uniformly random bit string of length $N$. Thus to classically simulate the output of the circuit we first classically choose a bit string $y_0$ uniformly at random and consider the state
\[ \ket{\phi_{y_0}} = \frac{1}{\sqrt{2^M}} \sum_x e^{if(x,y_0)} \ket{x}. \]
Since $\ket{\phi_{y_0}}$ is a state of only $O(\log n)$ qubits (i.e. $\poly (n)$ dimensions) we can classically strongly simulate the results of further gates and measurements on it, in $\poly (n)$ time by direct calculation, giving overall an exact weak classical simulation of the original circuit family's output.\, $\Box$

The methods developed in \cite{maarten} may also be used to readily provide a weak classical simulation up to additive polynomial error for the families in the above theorem.

Shepherd in \cite{newdan} gives a series of further classical simulation properties of IQP circuits. In particular it is shown there that the distributions $P_w$ in the above theorem are not only exactly weakly simulable, but even more, they are classically strongly simulable, if all the gates are restricted to have diagonal entries of only integer powers of $e^{i\pi /8}$ (which suffices, as we have noted, to obtain the conclusions of theorems \ref{postthm} and \ref{mainthm}).

As introduced in \cite{shbr}, we may consider a more general notion of an IQP assisted classical computation, than just the single use of the output of a uniform family of IQP circuits. Let $\ciqp$ denote an oracle, which if given a description $C$ of an IQP circuit, will obligingly return (in one computational step) a sample of C's output distribution. Then we may consider complexity classes such as BPP$^\ciqp$, defined as the class of languages decided with bounded error by a classical probabilistic polynomial-time computation where in addition to the usual classical steps, the computation may query the oracle with IQP circuit descriptions that have been produced as intermediate results along the way. Since any IQP circuit is a particular kind of quantum circuit, it is easy to see that BPP$^\ciqp \subseteq$ BQP, and theorem \ref{simlogthm} shows that BPP$^{\ciqp [\log n]} =$ BPP, where BPP$^{\ciqp [\log n]}$ denotes that the oracle is queried only with IQP circuits having at most $O(\log n)$ output lines.

\section{Some further remarks}\label{endisnigh}
It is interesting to note that the methods used to prove our principal results in theorem \ref{mainthm} and corollary \ref{maincor} may be applied to other classes of circuits. The only feature of IQP that we needed was the result of theorem \ref{postthm}, that post-selection boosts its power to PP. Thus the evidence of hardness of classical simulation provided by corollary \ref{maincor} would apply to any class of circuits that similarly goes to PP under post-selection. For example, the constructions in \cite{terdiv, fenetc} (exploiting the notion of gate teleportation \cite{gc}) imply that the power of quantum circuits of depth 4 (i.e. 3 layers of unitary gates followed by a layer of measurements) with post-selection includes BQP and hence also post-BQP $=$ PP, while quantum circuits of depth 3 are known to be always strongly classically simulable. More formally \cite{fenetc} introducing the class BQNC$^0$ of languages decided with bounded error by uniform families of constant depth circuits, we have post-BQNC$^0=$ PP and the conclusion of our corollary \ref{maincor} then applies to QNC$^0$ (constant depth quantum circuits) replacing IQP.

\bigbreak
\noindent
{\large\bf Acknowledgements.} MJB acknowledges the support of COQUIT, and in part, of the National Science Foundation under Grant No. PHY05-51164 while visiting the KITP. RJ was supported by the UK EPSRC QIPIRC and the EC network QICS. DJS was funded by CESG. We thank Aram Harrow and Ashley Montanaro for interesting discussions.

\end{document}